\documentclass[aps,prl,twocolumn]{revtex4}
\usepackage{graphicx}
\usepackage{color}

\newcommand{\be}{\begin{equation}}
\newcommand{\ee}{\end{equation}}
\newcommand{\bea}{\begin{eqnarray}}
\newcommand{\eea}{\end{eqnarray}}
\newcommand{\oo}[1]{{\cal O}\left( #1 \right)}    

\begin{document}

\title{Optimizing the vertebrate vestibular semicircular canal:  could we balance any better?}

\author{Todd M. Squires}
\affiliation{Departments of Physics and Applied and Computational Mathematics, California Institute of Technology 114-36, Pasadena, CA 91125}

\date{\today}

\begin{abstract}
The fluid-filled semicircular canals (SCCs) of the vestibular system are used by all vertebrates to sense angular rotation.  Despite masses spanning seven decades, all mammalian SCCs are nearly the same size.  We propose that the SCC represents a sensory organ that evolution has `optimally designed'.   Four geometric parameters are used to characterize the SCC, and `building materials' of given physical properties are assumed.  Identifying physical and physiological constraints on SCC operation, we find that the most sensitive SCC has dimensions consistent with available data.  
\end{abstract}
\maketitle

Simple physical principles and scaling arguments have been remarkably effective in understanding organismic and evolutionary properties of the animal kingdom \cite{vogel88}.  Sensory organs, which transduce physical stimuli into neural signals, are particularly well-suited for physical analysis. The biophysics of the cochlea, for example, has been the focus of much recent study \cite{gummer03}.  

Here we focus on a different inner-ear organ:  the fluid-filled semicircular canals (SCCs) that each of the roughly 45,000 vertebrate species employs to mechanically sense rotation \cite{bialek85,parker80}.  Rotation sensation is obviously imperative for balance, but also plays an equally important role in vision.  The neural output of the SCCs feeds directly to the oculomotor system, causing a reflexive motion of the eyes that compensates for head motion.  This allows you to read this article even while shaking your head, a task which would be much more difficult if the article itself were shaken.

\begin{figure}
\begin{center}
\centerline{
\includegraphics[width=2.5in]{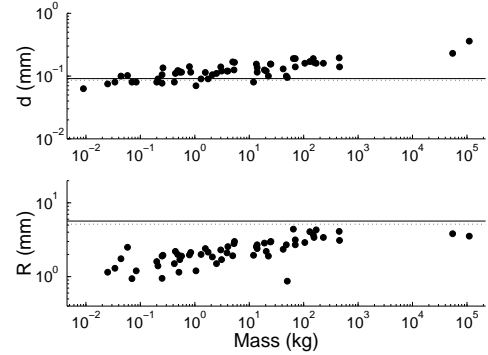}}
\caption{\label{fig:jonesspells} Measured values of $R$ and $d$ for mammals \cite{jones63,muller99}.  Despite seven decades of mass variation, variations in $R$ and $d$ are limited to about one decade.  Solid (dashed) horizontal lines reflect `optimal' (`tolerant') SCC dimensions (\ref{eq:rcopt}-\ref{eq:tolerant}).  Rough measurements of $c$ and $t$ for mammals \cite{gray07} reveal $t\sim 0.15-0.44$ mm and $c\sim 0.17-0.5$ mm (c.f. optimal $c=0.12,t=0.12$.)}
\end{center}
\end{figure}

Previously, we studied `top-shelf vertigo', a mechanical disorder of the human SCC \cite{squires04}.   Here, we address a rather astonishing feature of the SCCs:  the SCCs of every mammal, from mice to whales, are essentially {\sl the same size} (Fig. \ref{fig:jonesspells}) \cite{jones63,muller99}.  Mammals span seven decades in mass and almost three in length, yet SCC dimensions are restricted to less than one.  In fact, the SCC of the human fetus reaches its full adult size by the fourteenth week of pregnancy \cite{hoshino82}.  Fish, reptile, amphibian, and bird SCCs are of similar size \cite{jones63}, the one apparent exception being sharks (discussed below) \cite{muller99}.    In exploring reasons for the constancy of cell size, Vogel writes, ``...anything biological that doesn't vary in size ought to strike us as noteworthy'' \cite{vogel88}.  Similarly, SCC non-scaling suggests that these particular dimensions may be special to, and perhaps optimal for, their function. 

The idea that biological structures are somehow `optimal' has recently found rather dramatic support.  Microcavities in the brittle star skeleton act as perfect lenses \cite{aizenberg01}, and sea sponges develop single-mode optical fibers that rival current technology \cite{sundar03}.  The human visual system operates at the single-photon level \cite{baylor79}, and the auditory system is limited by thermal noise \cite{denk89}.  In fact, evolution can be viewed as a gradient search seeking to optimize `fitness'; however, the utility of this approach is often limited by the difficulty of defining `fitness'.  By contrast, the semicircular canals have a well-defined purpose, and reasonable measures of their quality can be proposed.

\begin{figure}
\begin{center}
\centerline{
\input{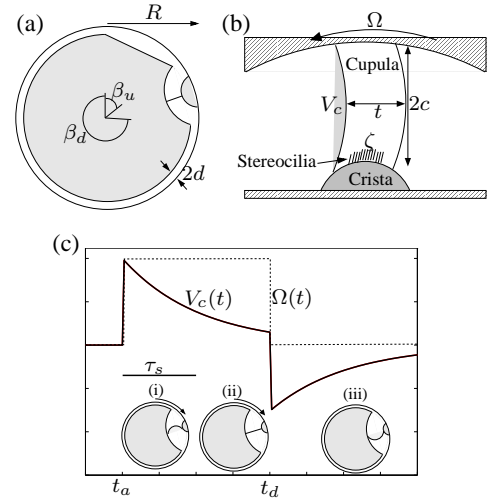}}
\caption{\label{fig:schematic} (a)  The semicircular canal is a torus of major radius $R$, with a long narrow duct of radius $d$ spanning an angle $\beta_d$.  (b)  The ampulla contains an elastic cupula of thickness $t$ and radius $c$, which distorts under SCC rotation, causing embedded hair cell stereocilia to deflect a distance $\zeta$ and trigger a neural signal.  Measured values (in mm) for humans are $\{R_h,c_h,d_h,t_h\}=\{2.8,0.44,0.19,0.31\}$ \cite{gray07}. (c)  SCC operation:  (i) Angular velocity is impulsively accelerated to $\Omega_0$ at $t_a$ and the cupula distorts by $V_c\propto\Omega_0$.  (ii) Under sustained rotation, the cupula relaxes and `forgets' $\Omega_0$.  (iii) Upon decelerating to rest at $t_d$, fluid inertia causes a (negative) cupular displacement, eliciting dizziness.}
\end{center}
\end{figure}

Herein, we suggest that the `universal' size of the SCCs can be understood in terms of a system which has been `optimally designed' by evolution.  Others have examined the relation between SCC sensitivity and geometry from a purely fluid standpoint \cite{jones63,muller99}, but did not address the elastic membrane and sensory hair cells that complete the transduction process, and which depend on SCC dimensions.  Our approach is as follows:  we start with SCC `blueprints' without dimensions (Fig. \ref{fig:schematic}), and assume basic building materials (solid walls, fluid endolymph, elastic material, and hair cells) to be given.  Identifying physically and physiologically reasonable constraints, we find a unique set of well-constrained and robust `optimal' dimensions that maximize SCC sensitivity.  Moreover, they are consistent with measured data (Fig. \ref{fig:jonesspells}).

We begin with a description of basic SCC structure (Fig. \ref{fig:schematic}).  Each ear contains three mutually orthogonal SCCs to span the three rotation axes.  Each SCC is a hollow torus of major radius $R$ that consists of a narrow {\sl duct} of radius $d$ and a bulbous {\sl ampulla} of radius $c$, all filled with water-like {\sl endolymph} of density $\rho$ and viscosity $\mu$.  The {\sl cupula}, a mucus membrane of thickness $t$ and radius $c$, completely spans the ampulla to block fluid flow (Fig. \ref{fig:schematic}b) \cite{bialek85}.  Along one wall sits sensory {\sl hair cells}, from which bundles of {\sl stereocilia} (of length $\lambda_H\approx 50\mu$m \cite{hillman72}) project into the cupula.  Within each bundle, stereociliar tips are linked by filaments that act as `gating springs', opening or closing ion channels when stereociliar tips are displaced \cite{markin95}.  Thus cupular deformations displace stereociliar tips, which cause the hair cells to fire.

Basic SCC function is shown in Fig. \ref{fig:schematic}c.  An impulsive angular acceleration $\alpha(t) = \Omega_0 \delta(t-t_a)$, viewed in the rest frame of the canal, gives rise to ballistic endolymph motion with initial angular velocity $\Omega_0$.  Viscous resistance from the walls cuts off this inertial motion after a time $\tau_f$ ($\sim5$ ms in humans), during which a volume $V_c\propto\Omega_0$ of fluid passes through the canal.  The cupula distorts to displace the same volume (Fig. \ref{fig:schematic}c(i)), which deflects embedded stereocilia and triggers a corresponding neural signal.
The elastic restoring force of the distorted cupula drives fluid back through the narrow duct.  Viscous resistance from duct walls slows cupular relaxation to a time scale $\tau_s$ (4-7 s in humans and primates \cite{dai99,goldberg71a}), so that  $V_c$ encodes angular velocity $\Omega$ for $t \lesssim \tau_s$.  Under sustained rotation, however, the cupula relaxes (Fig. \ref{fig:schematic}c(ii)) and `forgets' the constant rotation rate $\Omega_0$.  Upon stopping, the sudden deceleration leads to a cupular displacement in the opposite direction (Fig. \ref{fig:schematic}c(iii)), causing dizziness (the false sensation of rotation).  The shorter $\tau_s$, the more quickly an organism becomes dizzy.

Simple arguments give scaling relations for SCC processes.  Fluid `knows' that walls are accelerating only after vorticity created at the walls diffuses (with diffusivity $\nu=\mu/\rho$) to the duct center, giving an inertial time $\tau_f \sim d^2/\nu$.  During $\tau_f$, fluid moves ballistically with velocity $\Omega_0 R$, so that a fluid volume 
$V_c \sim (\pi d^2) \Omega_0 R \tau_f$ travels through the duct.  Once inertia is cut off, the cupula relaxes quasistatically, as the elastic restoring pressure $\Delta P = K V_c$ drives a Poiseuille flow $\dot{V}_c \sim \Delta P d^4 / \mu \beta_d R$, assuming viscous resistance to be dominated in the narrow duct.  An ODE for $V_c$ results, whose solutions decay exponentially on a time scale $\tau_s \sim \mu \beta_d R/K d^4.$

Prefactors were obtained by solving for the time-dependent flow due to an impulsive acceleration, giving
\be
V_c =  \frac{4\pi d^4 R\alpha}{\lambda_0^4\nu} \Omega_0,\,\,\,\,\tau_f = \frac{d^2}{\nu \lambda_0^2},\,\,\,\,\tau_s = \frac{8 \mu \beta_d R}{K \pi d^4},
\label{eq:maxdisp}
\ee
were $\lambda_0 \approx 2.4$ is the first zero of the Bessel function $J_0$, and $\alpha\approx 1.3$ is a geometric factor \cite{vanbuskirk76}.

The final physical ingredient involves elastic deformations of the cupula, which we treat as a clamped plate of modulus $E_0=E'(1-\sigma^2)$ and Poisson ratio $\sigma$ \cite{rabbitt92}.  A plate of bending rigidity $D_0 = E't^3/12$ obeys $D_0 \nabla^4 w = \Delta P$, giving a deformed profile $w = w_{{\rm max}} [1-(r/c)^2]^2,$ where $w_{{\rm max}} = \Delta P c^4/64 D_0$ \cite{landauelasticity}.  An integration gives the cupular stiffness $K$ relating $\Delta P_c$ to $V_c$,
\be
K = E' \frac{16}{\pi} \frac{t^3}{c^6},
\label{eq:defk}
\ee
whose value $K=4.6$ GPa/m$^3$ is derived from measured values of $\tau_s\approx 4$ s and (\ref{eq:maxdisp}).  As a check, this implies $E' \approx 2$ dyne/cm$^2$, which is consistent with measurements for mucus \cite{powell74}.  

Having described the mechanical response, we now connect cupular displacement $w$ to stereociliar tip displacement $\zeta$ (and thus to a neural firing rate $f$).  Hair cells have a high resting discharge rate $f_r$ and are bi-directionally sensitive, so that deflections in either direction can be sensed quickly \cite{bialek85}.  The gating springs that link stereociliar tips cause ion channels to open or close as the tips are deflected, increasing or decreasing $f$ linearly with $\zeta$ for small deflections \cite{markin95}.  Since the tips are embedded in the cupula, their deflection $\zeta$ varies with $w$ via
\be
\zeta \approx \frac{4\lambda_H^2}{c^2}w_{{\rm max}} = \frac{12\lambda_H^2}{\pi c^4}V_c\,\,\,{\rm for}\,\lambda_H \ll c. 
\label{eq:zetav}
\ee

Measurements on 5-10 $\mu$m utricular and cochlear hair cells indicate a threshold displacement $\zeta_{\rm min}\sim \oo{{\rm nm}}$ for neural activity, and gating spring models predict $\zeta_{\rm min}$ to vary with hair cell length \cite{markin95}.  Because SCC hair cells are about 10 times longer, one expects $\zeta_{\rm min}\sim\oo{10\,{\rm nm}}$.  A similar value for $\zeta_{\rm min}$ can be obtained from other experiments under the assumption that (order of magnitude) hair cell properties are conserved across species.  Physiological measurements indicate a lower limit for sensation in humans, giving a sensation threshold $\Omega_{\rm min} \approx 2^\circ$/s \cite{oman72}.   
In squirrel monkeys, the relationship between $\Omega$ (or $\zeta$) and firing rate was measured to be linear: $\Delta f \approx \alpha \Omega$, where $\alpha = 0.24/^\circ$ \cite{goldberg71a}.  The firing rate has natural limits:  $f$ can not be negative, and fires at a maximum (injured) rate ($\sim$400 Hz). Saturating deflections $\zeta_{\rm sat}$ occur for $\Delta f \sim f_r\approx 100$ Hz, implying a `saturating' rotation $\Omega_{\rm sat} = f_r/\alpha \approx 420^\circ$/s.  Using (\ref{eq:maxdisp}) and (\ref{eq:zetav}), we find a dynamic range of stereocilia deflections
\be
\zeta_{\rm min} \approx 20\,{\rm nm}, \,\,\,\zeta_{\rm sat} \approx 4\,\mu {\rm m},
\label{eq:zetacrit}
\ee
and note that $\zeta_{\rm min}$ thus obtained is consistent with predictions of the gating spring model.

The sensitivity $S$ of an arbitrary SCC follows from (\ref{eq:maxdisp}) and (\ref{eq:zetav}), with minimum detectable rotation $\Omega_{{\rm min}}$,
\be
\Omega_{{\rm min}} = \frac{\lambda_0^4 \mu}{48 \lambda_H^2 \alpha}\frac{c^4}{d^4 R} \zeta_{{\rm min}}\equiv S^{-1} \zeta_{{\rm min}}.
\label{eq:sensitivity}
\ee

We have now characterized the mechanical-neural signal transduction process, and can examine the effect of varying SCC dimensions.  We scale SCC variables by their human values, denote scaled variables with hats, and pose the central question:  Given `blueprints' (Fig. \ref{fig:schematic}) and basic building materials ($\mu,\rho,E',\lambda_H, \zeta_{{\rm min}}$ and $\zeta_{{\rm sat}})$, what dimensions $\{\hat{R},\hat{c},\hat{d},\hat{t}\}$ maximize SCC sensitivity $\hat{S}=\hat{d}^4 \hat{R}/\hat{c}^4$?

Certain constraints, however, must be obeyed for a physically- and physiologically-viable SCC.
First, thermal fluctuations of the cupula must be small enough to escape detection, or else balance and vision would be impaired.  The probability of a cupular displacement fluctuation $V_c$ is given by $P(V_c)\sim \exp(-K V_c^2/2 k_B T)$.  We use (\ref{eq:defk}) and (\ref{eq:zetav}) to express $P(V_c)$ in terms of $\zeta$,  giving $P(\zeta)=2 \sqrt{\sigma/\pi} \exp(-\sigma \zeta^2),$ where $\sigma=\pi E' c^2 t^3/18 k_B T \lambda_H^4$.  The probability of sensing a thermal fluctuation is given by $P(\zeta > \zeta_{{\rm min}}) = 1-{\rm erf}(\sigma^{1/2} \zeta_{{\rm min}})$.  Requiring $P(\zeta>\zeta_{{\rm min}})$ to be smaller than some value $\epsilon = 10^{-5}$ gives
\be
\hat{c}^{2/3} \hat{t}>\left(9.7\frac{18\lambda_h^4 k_B T}{\pi E' \zeta_{{\rm min}}^2 c_h^2 t_h^3 }\right)^{1/3}
\equiv A =0.15.
\label{eq:consta}
\ee
Note that $A$ is rather insensitive to $\epsilon$: $A(10^{-6}) = 0.16$ and $A(10^{-4})=0.14$.  

Second, the cupula should encode angular {\sl velocity} for as long as possible.  The smaller $\tau_s$, the sooner angular velocity is `forgotten' and dizziness ensues.  Requiring $\tau_s$ to be larger than some minimum time $\tau_s^m$ gives
\be
\frac{\hat{R} \hat{c}^6}{\hat{d}^4 \hat{t}^3} > \frac{2 E' \tau_s^m}{\mu\beta_d}\frac{t_h^3 d_h^4}{R_h c_h^6} \equiv B=0.25,
\label{eq:constb}
\ee
and we take $\tau_s^m = 1$ s as a basic estimate.

Third, the maximum cupular deflection $w_{{\rm max}}$ for physiological rotations should be small (compared to $t$), to avoid damaging the cupula and for a linear response.  The saturating rotation $\Omega_{\rm sat}$ is one example, giving 
\be
\frac{\hat{c}^2}{\hat{t}}< \frac{4\lambda_h^2}{\zeta_{\rm sat}}\frac{t_h}{c_h^2} \equiv C=4.1.\label{eq:constc}
\ee
Rotations $\Omega_{\rm max}$ which saturate the neural signal, but are still physiologically relevant, also impose a constraint.  As a basic estimate, we take $\Omega_{\rm max} = 2 \pi$ rad/s, which gives
\be
 \frac{\hat{d}^4 \hat{R}}{\hat{c}^2 \hat{t}} < \frac{\lambda_0^4 \nu}{12 \alpha \Omega_{\rm max}}\frac{c_h^2 t_h}{d_h^4 R_h} \equiv D=4.5.\label{eq:constd}
\ee

Finally, the ampulla must fit in the canal,
\be
\hat{R}/\hat{c} > 2 c_h/R_h \equiv E = .32, \label{eq:conste}
\ee
and the cupula should be plate-like for uniform $\zeta$,
\be
\hat{c}/\hat{t}> t_h/c_h\equiv F = 0.70. \label{eq:constf}
\ee

\begin{figure}
\begin{center}
\centerline{
\input{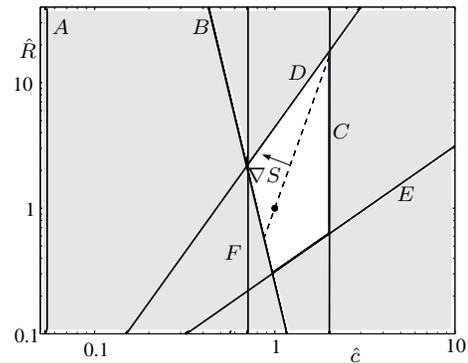}}
\caption{\label{fig:constraint} A 2D slice of 4D parameter space showing the constrained space available for viable SCC design.  Constraints (\ref{eq:consta}-\ref{eq:constf}) are labeled A-F.  Also plotted is the sensitivity gradient and, for reference, dimensions of the human SCC.  The `optimal' SCC is found at the vertex of constraints A, B, D, and F. }
\end{center}
\end{figure}

Eqs. (\ref{eq:sensitivity}-\ref{eq:constf}) represent a nonlinear multidimensional optimization, but an equivalent linear optimization is obtained by taking the logarithm of each equation.  As in linear programming, extrema are found at vertices of the constraint equations (or $-\infty$, since logs can be negative).  Here, the most sensitive SCC geometry is found at vertices $A$,$B$,$D$, and $F$, giving dimensions
\be
\hat{R}=\frac{(BD)^{1/2}}{F^2}=2.1,\,\hat{c}=(AF)^{3/5}=0.26,\label{eq:rcopt}
\ee
\be
\hat{d}=\frac{A^{9/20}D^{1/8}F^{7/10}}{B^{1/8}}=0.49,\,\hat{t}=\frac{A^{3/5}}{F^{2/5}}=0.38,\label{eq:dtopt}
\ee
and a sensitivity $\hat{S}$,
\be
\hat{S} = DA^{-3/5} F^{-8/5}\approx 24,\label{eq:sensopt}
\ee
that is about 24 times greater than the human SCC.  As seen in Fig. \ref{fig:jonesspells}, these dimensions compare quite favorably with available data---particularly since such an optimum need not exist in the first place, and even if it did, its dimensions could have differed from those found in nature by many orders of magnitude. 

Two parameters ($\Omega_{{\rm max}}$ and $\tau_s^m$) were chosen somewhat arbitrarily, and it is natural to question how these choices affect the above results.  Only $\hat{R}$ and $\hat{d}$ depend on these choices, and weakly at that: $\hat{R}\sim (\tau_s^m/\Omega_{{\rm max}})^{1/2}$ and $\hat{d}\sim(\tau_s^m\Omega_{{\rm max}})^{-1/8}$.  For $\hat{R}$ to change by an order of magnitude, $\tau_s^m/\Omega_{{\rm max}}$ must be off by two.  This insensitivity to the crude choices made above reinforces the robustness of this result, and implies that existing inter-species variations in $\tau_s^m$ and $\Omega_{{\rm max}}$ should lead to only slight differences in optimal SCC dimensions.

While the above analysis assumes perfect `machine tools,' nature's machinery has a rather larger tolerance: variations in human SCC dimension are of order 10\% \cite{curthoys87}.  To allow for imperfect `machining', we find the most sensitive `tolerant' SCC such that 10\% variations in any dimension satisfy constraints $A-F$, giving
\be
\hat{R} = 1.9,\,\hat{c}=0.33,\,\hat{d}=0.46,\,\hat{t}=0.38,\label{eq:tolerant}
\ee
with a relative sensitivity $\hat{S} \approx 7$.  SCC sensitivity variations
due to size variations range from about 0.4 to 2.5 times that of the `average' canal.  


No constraint on measurement time $\tau_f$ was introduced, although it plays an important role in vision: to be `watched', an image must be kept within one retinal fovea -- about $1^\circ$ in humans \cite{bialek85}.  In the time $\tau_f$ following an $\Omega_0$ impulse,  the eye/fovea lags the image by $\Delta \theta \sim \Omega_0 \tau_f$, so that the maximum rotation allowing a fixed gaze is $\Omega_v \sim 1^\circ/\tau_f \sim 200^\circ$/s.  Larger creatures typically undergo slower rotations, allowing larger $\tau_f$ (and thus $d$).  Furthermore, rapidly-flying birds have smaller $\hat{d}$ than similarly-sized earthbound ones \cite{bialek85}, which may allow them to hold objects in view more easily during rapid maneuvers. 

Although we have explicitly discussed mammals, the same SCC dimensions are found in all vertebrates, with a few notable exceptions.  Head size is an obvious and unavoidable constraint for exceedingly small fish larvae, whose SCCs start small and grow with the fish \cite{muller99}.  As a group, sharks--even small ones--have abnormally large SCCs ($R \sim 40$ mm) \cite{muller99}.  Since sharks broke rather early from other vertebrates in evolution, a possible explanation could involve different `building materials'.  
 
In summary, we have argued that evolution has converged on an `optimal' design for a maximally sensitive rotation detector.  The optimal SCC is well-constrained, with little room for variations, and falls within a factor of three of available data.  
The optimum itself is robust and depends on basic mechanics and established hair cell properties; the `calibrating' assumptions made herein are self-consistent, but not essential for this central result.  

I am grateful to Howard Stone for introducing me to the vestibular system, and for many thoughtful comments and discussions.  Conversations with Bill Bialek about foveation, with Michael Brenner about optimization, and with Eric Lauga about this article are gratefully acknowledged.

\end{document}